\documentclass[]{iopart}
\usepackage{iopams}
\usepackage{setstack}
\usepackage[]{graphicx}
\usepackage{color}
\usepackage[latin1]{inputenc}

\newcommand{\NA}{{N_{\rm A}}}
\newcommand{\eSi}{{^{28}{\rm Si}}}

\begin{document}

\title[]{Lattice strain at c-Si surfaces: a density functional theory calculation}
\author{C Melis$^1$, L Colombo$^1$ and G Mana$^2$}
\address{$^1$Department of Physics, University of Cagliari, Cittadella Universitaria, 09042 Monserrato (Ca), Italy}
\address{$^2$INRIM - Istituto Nazionale di Ricerca Metrologica, Str.\ delle Cacce 91, 10135 Torino, Italy}
\ead{claudio.melis@dsf.unica.it}

\begin{abstract}
The measurement of the Avogadro constant by counting Si atoms is based on the assumption that Si balls of about 94 mm diameter have a perfect crystal structure up to the outermost atom layers. This not the case because of the surface relaxation and reconstruction, the possible presence of an amorphous layer, and the oxidation process due to the interaction with the ambient. This paper gives the results of  density functional calculations of the strain components orthogonal to crystal surface in a number of configurations likely found in real samples.
\end{abstract}

\submitto{Metrologia}



\section{Introduction}
International efforts are on going on accurate determinations of the Planck, $h$, and Avogadro, $\NA$, constants \cite{Bettin:2013}. They are motivated by the possibility to replace the definition of the unit of mass by a definition based on a conventional value of the Planck constant, $\NA$ and $h$ being linked by the molar Planck constant, $\NA h$, which can be accurately measured \cite{Massa:2012}.

The most accurate way to determine $\NA$ is by counting the number of atoms in a Si ball highly enriched with $\eSi$ \cite{Andreas1:2011,Andreas2:2011}. The count is carried out by dividing the molar volume, $VM/m$ --  where the symbols indicate the volume, molar mass, and mass of the ball, by the volume occupied by an atom in a perfect face centered cubic crystal with a 2-atom basis, $a_0^3/8$ -- where $a_0$ is the lattice parameter. The uncertainty associated to the presently most accurate determination is about $3 \times 10^{-8} \NA$. In order to achieve this accuracy level, the ball volume is measured to within a $2\times 10^{-8}V$ uncertainty. This uncertainty corresponds to a determination of the about 47 mm mean ball-radius to within an uncertainty of 0.3 nm, that is, given or taken one atomic layer.

Atom counting is based on the assumption that the ball has a perfect crystal structure up to its outermost atom layers; this not the case because of surface reconstruction, surface stress and strain, and the possible presence of amorphous or oxide layers. Surface effects are of interest in micro- and nano-mechanics and, therefore, they have been the subject matter of extensive investigations; a few relevant to our problem are in \cite{Hara:2005,Delph:2008,Qi:2009a,Qi:2009b}.

When a silicon crystal is cleaved to form a surface, an overall surface relaxation and/or surface reconstructions take place. The reconstructions are due to the presence of dangling bonds (created upon the surface cleavage) which tend to rebind forming a network other than the bulk like one. Relaxations and reconstruction stress and strain the atom layers nearest to the surface. Such surface stress and strain may in turn have a twofold effect on the $\NA$ measurement. Firstly, the surface stress may make the lattice parameter of a Si ball different from the value measured in the crystals of an x-ray interferometer, which are only about 1 mm thick \cite{Quagliotti:2013}. Secondly, the surface strain may sink or raise the ball surface and make the measured volume smaller or greater than the volume of the unstrained crystal.

This paper focuses on the second issue. We report about density functional theory calculations of surface strain with atomistic resolution and aimed at excluding or quantifying systematic surface contributions to the measured $\NA$ value. Calculations were carried out by using the Quantum Espresso computer package \cite{QE:2009}.

Section 2 starts by outlying the way $\NA$ is measured. Section 3 outlines the concepts of the density functional theory and the tunings made to achieve the maximum accuracy in the calculation of the perfect-crystal lattice parameter. Next, in section 4, we give the numerical estimates of the interatomic distances in the outmost lattice planes. Eventually, section 5 estimates the contribution of the sphere surface to the uncertainty of the $\NA$ measured-value.

\section{The $\NA$ measurement}
The value of the Avogadro constant is obtained from measurements of the molar volume, $VM/m$, and lattice parameter, $a_0$, of an ideally perfect and chemically pure silicon mono-crystal. In a formula,
\begin{equation}\label{NA}
 \NA = \frac{8MV}{a_0^3 m} ,
\end{equation}
where $m$ and $V$ are the crystal mass and volume, $M$ is the mean molar mass, and 8 is the number of atoms in the cubic unit cell. To make the kilogram redefinition possible, the targeted accuracy level is $2\times 10^{-8} \NA$.

From (\ref{NA}), it follows that the $\NA$ determination requires the measurement of i) the lattice parameter -- by combined x-ray and optical interferometry \cite{Massa:2011}, ii) the amount of substance fraction of the three Si isotopes and, then, the molar mass -- by absolute mass-spectrometry \cite{Pramann:2011}, and iii) the crystal mass and volume \cite{Picard:2011,Kuramoto:2011,Bartl:2011}. Silicon crystals do contain chemical impurities as well as point  and extended native defects, which implies that the measured mass value does not correspond to that of an ideal Si crystal, and that the crystal lattice may be distorted. This means that crystals must be characterized both structurally and chemically, so that the appropriate corrections can be applied \cite{Fujimoto:2011,Massa:2011b,Zakel:2011}.

In order to carry out an accurate volume measurement, the crystal is shaped as a nearly perfect ball having about 94 mm diameter and whose surface deviates from a sphere no more than a few tens of nanometre. The ball volume is calculated by combining a survey of the surface topography with absolute diameter measurements simultaneously obtained by embedding the Si ball into a spherical optical-resonator \cite{Kuramoto:2011,Bartl:2011}.

In order to achieve the targeted $\NA$ uncertainty, the relative uncertainty of the mean-radius measurement must be reduced below $6\times 10^{-9}$, which corresponds to an absolute uncertainty of about 0.3 nm. Since (\ref{NA}) assumes an ideal Si crystal, the mass, thickness, and chemical composition of the oxide layer covering the ball are measured by optical and x-ray spectroscopy and reflectometry and subtracted from the measured mass and volume values \cite{Bush:2011}.

In addition, surface reconstruction strains the atom layers nearest to the surface and makes their lattice parameter different from that in the bulk. In order exclude a systematic effect on the volume measurement or to correct the measurement result, it is necessary to calculate the lattice parameter as a function of the distance from the surface. The next two sections describe the theoretical and numerical tools used to carry out this calculation and the results obtained.

\section{Density functional theory}\label{DFT}
Density functional theory (DFT) allows to solve the Schr\"odinger's equation for large and complex condensed matter systems (up to about $10^3$ atoms). DFT solves the electronic Schr\"odinger equation by reducing the quantum mechanical problem of a many-body interacting system to an equivalent problem for non-interacting particles. This is achieved by using as fundamental variable the electronic density instead of the many-body electronic wavefunction. Nowaday DFT is a well established tool for the study of the properties of many-body systems without using empirical parameters \cite{DFT}.

The theoretical base of  DFT is the Hohenberg and Kohn theorem \cite{Hohenberg} which considers an electronic system subject to the external potential $V^{\rm ext}({\bf r})$. This theorem states that i) the ground state density of the many-electron system uniquely determines  the external potential, modulo a constant, and ii) the ground state energy is the minimum of the total energy with respect to the electronic density $n({\bf r})$.

By considering a set of Hamiltonians that have the same kinetic energy and electron-electron operator but different external potentials, their ground states will have different densities. The external potential is thus a functional of the ground-state density. Once the external potential is fixed, also the total energy
\begin{equation} \label{eqn:energia}
E_V[n]=F[n]+\int \rmd{\bf r} \, n({\bf r}) V^{\rm ext}({\bf r}) ,
\end{equation}
will be a functional of $n({\bf r})$ where $F[n]$ is a universal functional of the density defined by the kinetic energy and electron-electron interaction. The minimum is obtained when $n({\bf r})$ is the ground state density.

The functional $F[n]$ is the most important part of $E_V[n]$, but there is no analytic expression for it and it is not easy to calculate. Kohn and Sham \cite{Sham} proposed an approximate expression by considering an equivalent problem  of non interacting electrons. The core of their assumption is that, for each system of interacting electrons, a corresponding system of non-interacting particles exists, subject to the external potential $V_{KS}({\bf r})$ and having the same ground state density as the interacting system.

Accordingly, the Kohn-Sham functional can be written as \cite{Sham}
\begin{equation} \label{eqn:f}
F[n]=T_e[n]+U[n]+E_{\rm xc}[n] ,
\end{equation}
where $T_e[n]$ is the kinetic energy of noninteracting electrons with density $n({\bf r})$,
\begin{equation} \label{eqn:hartree}
U[n]= \frac{1}{2}\int \rmd{\bf r}\, \rmd{\bf r'}\, \frac{n({\bf r})n({\bf r'})}
{|\mathbf{r-r'}|}
\end{equation}
is the Hartree energy, i.e., the classical electrostatic energy corresponding to $n({\bf r})$, and $E_{\rm xc}[n]$ is the exchange and correlation energy.

By assuming that $E_{\rm xc}[n]$ is known, it is possible to treat the many-body system as a system of independent particles. The ground state of this system is obtained from the solutions of the single particle Kohn-Sham equations
\begin{equation}
\label{eqn:kohn-sham}
-\frac{1}{2}\nabla^2 \psi_i({\bf r})+V_{\rm KS}({\bf r})\psi_i({\bf r})  =
\varepsilon_i \psi_i({\bf r}) \nonumber \\
\end{equation}
where
\begin{equation}
V_{\rm KS}=V^{\rm ext}+\varphi({\bf r})+v_{\rm xc}({\bf r})
\end{equation}
is the Kohn-Sham potential, $\varphi({\bf r})$ is the classic electrostatic potential of a charge distribution $n({\bf r})$, and $v_{\rm xc}({\bf r})$ is the functional derivative of the exchange and correlation energy. The eigenfunctions $\psi_i({\bf r})$ (with the orthonormality condition $\int \rmd{\bf r} \, {\psi^*}_i({\bf r}) \psi_j({\bf r})=  \delta_{ij}$) are called Kohn and Sham orbitals. Since $V_{\rm KS}$ depends on $n({\bf r})$, the Kohn and Sham equations must be solved in a  self-consistent way.

Dealing with $E_{\rm xc}[n]$ is the most difficult task in the solutions of the Kohn-Sham equations. The Pauli exclusion principle imposes the antisymmetry of the many-electron wavefunction. This antisymmetrization produces a spatial separation between the electrons having the same spin and reduces the Coulomb energy of the system. This reduction is the exchange energy for which an exact description is only provided by the Hartree-Fock method; the difference between the energy of an electronic system and the Hartree-Fock energy is the correlation energy. It is extremely difficult to calculate the correlation energy of a complex system, although some attempts have been made by using quantum Monte Carlo simulations.

The most popular approximations for $E_{\rm xc}[n]$ are the  Local Density Approximation (LDA) and the  Generalised Gradient Approximation (GGA). LDA assumes that the local exchange-correlation energy $\varepsilon_{\rm xc}[n]$ is equal to that of a homogeneous electron gas having the same density as the electron gas at ${\bf r}$. It assumes that $\varepsilon_{\rm xc}[n]$ is purely local, ignoring the corrections due to the nearby inhomogeneities of the electron density. GGA uses the series expansion of the electron density. Generally, the expansion stops at the first derivative and $\varepsilon_{\rm xc}[n]$ is expressed as a function of $n({\bf r})$ and $|\nabla n({\bf r})|$. GGA approximations can be developed at different  levels of sophistication, including PBE \cite{PBE}, BLYP \cite{BLYP,BLYP1}, BP \cite{PBE,BP}, PW91 \cite{PW91}, and PBESOL \cite{PBESOL}

\section{Methods}
All the  calculations have been performed using the Quantum Espresso computer package \cite{QE:2009}, an integrated suite of Open-Source computer codes for electronic-structure calculations and materials modelling at the nanoscale based on density-functional theory, plane waves, and pseudopotentials. Preliminary calculations were focused to benchmark the fundamental parameters of the DFT calculations, namely: the exchange and correlation functional, the pseudopotential type, the number of plane waves, and the number of $k$-points. All benchmark calculations have been performed on bulk crystalline silicon (c-Si) since its surface relaxations are the object of our analysis and bulk c-Si is, therefore, the natural choice as prototypical system to be used for assessing the reliability of DFT against experiments.

The first goal was to select the combination of the parameters giving the most accurate description of the c-Si lattice constant as compared to the experimental values. As far as concerns the exchange-correlation functional, we considered three different options: the Local Density Approximation (LDA), the General gradient approximation (PBE), and the General gradient approximation (PBESOL).

As for the pseudopotential type, we considered: the Goedecker-Hartwigsen-Hutter-Teter pseudopotentials (HGH) \cite{HGH}, the Martin-Troullier pseudopotentials (MT) \cite{MP}, and the plane augmented wave pseudopotentials (PAW) \cite{PAW}. It must be noted that HGH and MT pseudopotentials are norm-conserving, while PAW is ultrasoft.

Finally, as for the $k$-points, we considered meshes of points ranging from 2$\times$2$\times$2 $k$-points up to 16$\times$16$\times$16 $k$-points distributed according to the Monkhorst-Pack algorithm \cite{MP} within the first Brillouin zone of the reciprocal lattice. Finally, we considered kinetic energy cutoffs for the one-electron wavefunctions ranging from 20 Rydberg to 200 Rydberg.

The Table \ref{T0} shows the calculated lattice parameters values as a function of the exchange-correlation functionals and pseudopotential types. All the calculated values were obtained using a 16$\times$16$\times$16 $k$-point mesh distributed according to the Monkhorst-Pack algorithm and a kinetic energy cutoff for the one-electron wavefunctions of 200 Rydberg. The best agreement with the experimental value is obtained by using the novel PBESOL exchange-correlation functional, which is specifically designed to calculate the bulk properties of solids. The agreement is excellent, with a relative error as small as 1$\times 10^{-4}$ and shows that the calculation of the c-Si lattice constant is extremely reliable.

\begin{table}[t]
\caption{\label{T0}Calculated values of the lattice parameter, $a_0^{DFT}$, as a function of the exchange-correlation functionals (EX-C) and of the pseudopotential. Calculations were done by using $16\times 16\times 16$ $k$-points (distributed according to the Monkhorst-Pack algorithm within the first Brillouin zone of the reciprocal lattice) and a kinetic energy cutoff for the one electron wave functions of 200 Rydberg. The experimental value of the lattice parameter is $a_0^{\rm expt} = 543.102$ pm \cite{Massa:2011}.}
\begin{indented}
\item[]\begin{tabular}{llrr}
\br
EX-C &pseudopotential &$a_0^{DFT}$ / pm  &$(a_0^{DFT}-a_0^{expt})/a_0$ \\
\mr
LDA     &nc-HGH  &538.100 &$-9.2\times 10^{-3}$ \\
LDA     &nc-MT   &537.986 &$-9.4\times 10^{-3}$ \\
PBE     &nc-HGH  &546.054 &$+5.4\times 10^{-3}$ \\
PBE     &nc-MT   &546.445 &$+6.2\times 10^{-3}$ \\
PBE     &us-PAW  &546.612 &$+6.5\times 10^{-3}$ \\
PBESOL  &us-PAW  &543.041 &$-0.1\times 10^{-3}$ \\
\br
\end{tabular}
\end{indented}
\end{table}

Moreover, our preliminary benchmark calculations provided evidence that the fully converged value is obtained by considering a 4$\times$4$\times$4 $k$-point mesh and a plane-waves kinetic energy cutoff of 35 Rydberg. Such a combinations guaratees a reduced computational cost having the same overall accuracy as the 16$\times$16$\times$16 mesh combined with 200 Rydberg calculations. Based on these results, we performed all the subsequent calculations by using i) the exchange-correlation functional PBESOL, ii) the pseudopotential Ultrasoft PAW, iii) a $4\times 4\times 4$ $k$-points mesh, and iv) a 35 Rydberg plane-waves kinetic energy cutoff.

\section{Calculation of the surface strain}\label{CLS}
Our goal is to estimate the strain of c-Si surfaces upon structural relaxations and reconstructions. Since it is impossible to consider all the facet-orientations of a ball, we selected the most energetically stable, i.e., the (100), (110), and (111) ones, by taking into account both full hydrogenation as well as perfectly clean surfaces. For the (100) and (110) surfaces, reconstruction and, for the (110) surface, amorphization have been also considered. In a real laboratory sample the ball surface is covered by an oxide layer, less than 2 nm thick. In this work, however, in order to better focus our systematic survey, we only considered clean (or H-saturated) surfaces. The investigation of oxidised surfaces has been postponed to a subsequent work.

All calculations have been carried out on supercells having 20 layers of 8 Si atoms. The supercell dimensions are $(4215.210\times 1086.078\times 1086.078)$ \AA$^3$, $(5339.8763\times 767.975\times 1086.080)$ \AA$^3$, and $(3455.889\times 767.975\times 1330.172)$ \AA$^3$, respectively for the (100), (110), and (111) surfaces.

In all the calculations the topmost 16 Si layers were free to relax, while the bottom four layers were clamped at their ideal lattice position in order to simulate the semi-infinite bulk structure underlying the surface. Relaxations have been accounted for by force minimization, until forces on unconstrained atoms vanish within 0.005 eV/\AA. The hydrogenated supercells contains 160 Si atoms and 16 H atoms, while non-hydrogenated ones contains just 160 Si atoms.

\begin{figure}\centering
\includegraphics[width=7.5cm]{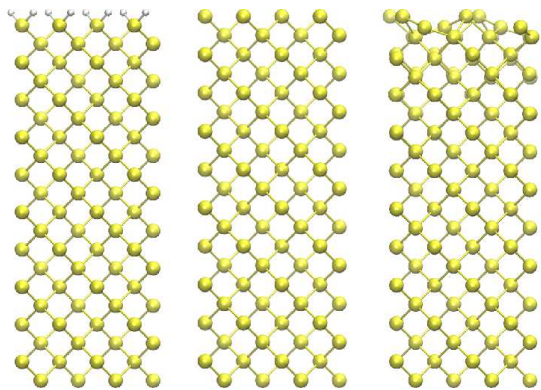}
\caption{Supercells used to calculate the out-of-plane strain of the Si(100) surface. Left: hydrogenated surface; center: non-reconstructed and non-saturated surface; right: reconstructed p(2$\times$1) antisymmetric surface.} \label{100-2x2-supercell}
\end{figure}

\begin{figure}\centering
\includegraphics[width=7.5cm]{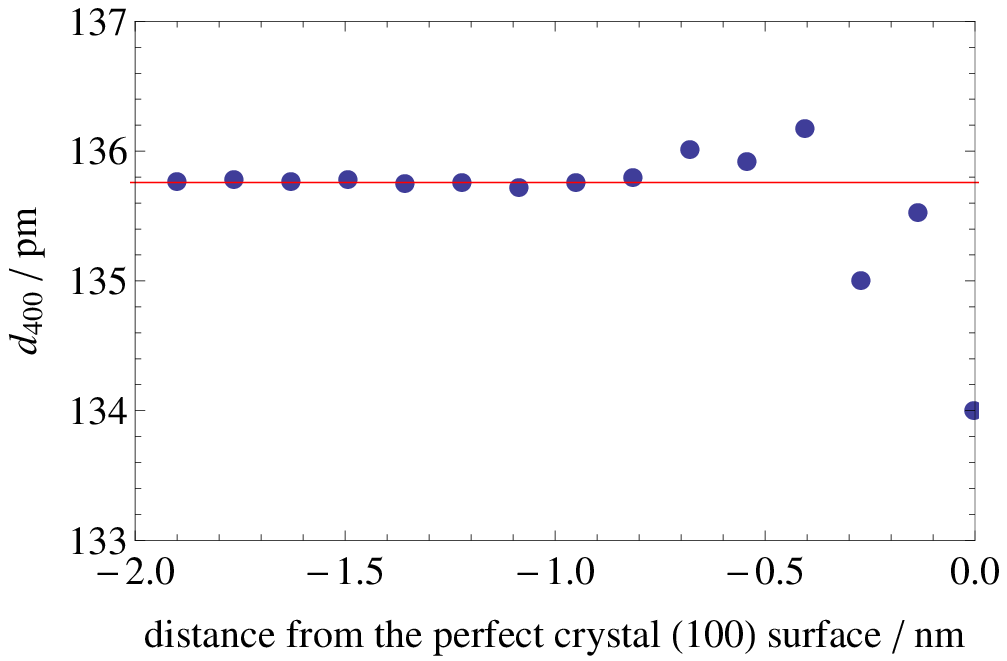}
\includegraphics[width=7.5cm]{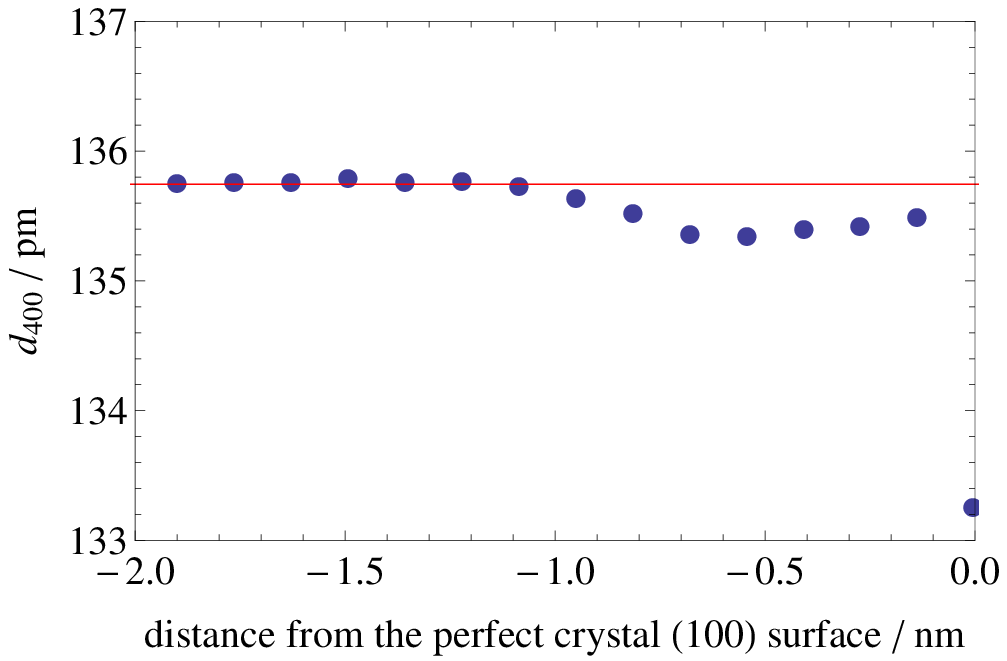}
\includegraphics[width=7.5cm]{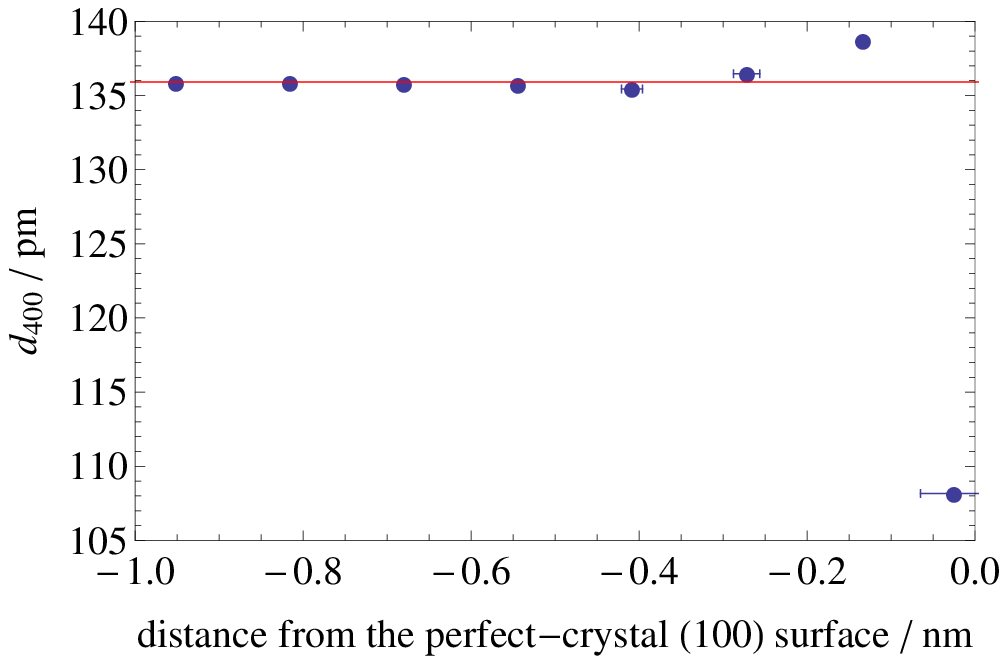}
\caption{Spacing $d_{400}$ of the \{400\} lattice planes as a function of the distance from the (100) surface of a perfect crystal. Top: non-reconstructed and non-saturated surface; middle: non reconstructed and hydrogenated surface; bottom: p(2$\times$2) reconstruction. The depth of each atom layer (blue dots) is calculated by sorting the supercell atoms by their distance from the surface and by taking the average of each subsequent subsets of 8 atoms. The error bars indicate the minimum and maximum depth of the atoms of each subset. The horizontal (red) lines are the perfect-crystal value of $d_{400}$.} \label{100}
\end{figure}

\subsection{(100) surface}
As far as concerns the (100) surface, we considered three different fully relaxed configurations, namely: the case of fully hydrogenated, non-hydrogenated and unreconstructed, and clean but reconstructed surface (Fig.\ \ref{100-2x2-supercell}). Several (100) reconstructions have been experimentally observed \cite{exp-100} and theoretically proposed \cite{theo-100}; they include the p(2$\times$1) symmetric and antisymmetric, p(2$\times$2), and c(4$\times$2) ones. Despite such a huge body of work, there is no consensus in the literature as to the nature of the lowest energy reconstruction. For this reason we studied a prototypical Si(100) reconstruction, namely the p(2$\times$2) antisymmetric configuration, which has been identified as the most energetically favoured by several theoretical works and it has been also experimentally observed. Figure \ref{100-2x2-supercell} shows the final supercell; this reconstruction is characterized by rows of alternating buckled dimers.

After full relaxation, we estimated the $d_{400}$ spacing of the \{400\} planes -- the densest among those parallel to the surface -- as a function of the distance from the surface of a perfect crystal. Figure \ref{100}-top shows the $d_{400}$ spacing for the non-reconstructed and non-saturated surface. We remark that the zero depth is assigned to the perfect crystal surface, while negative depth values correspond to the inner atomic planes. We can distinguish two main regions: i) a bulk-like region, below $-0.6$ nm, where the lattice spacing is not significantly different from its bulk value and ii) a surface region, above $-0.6$ nm, where the lattice spacing decreases, in particular, nearby the last layer of Si atoms. Such a decrease is quite small, about 1\%. To our understanding, the nicely thick bulk-like region below $-0.6$ nm from the surface implies that the supercell was thick enough to take any real relaxation into account.

In the case of  the hydrogenated (100) surface, $d_{400}$ does not show any significant variation with respect to the perfect crystal values; the maximum variation is as small as about 2\% and involves only the last layer of atoms (Fig.\ \ref{100}-middle). Figure \ref{100}-bottom shows $d_{400}$ as a function of the distance from the (100) surface of a perfect crystal for the p$(2\times 2)$ surface reconstruction. Also in this case we can distinguish a bulk-like region (below $-0.2$ nm) and a surface region (above $-0.2$ nm). However, contrary to the previous case, we observe large variations of the lattice spacing. The Si(100) p$(2\times 2)$ reconstruction strongly affects the lattice spacing of the last two Si layers; nearby the last Si layer it is as large as 21\%.

\subsection{(110) surface}
\begin{figure}\centering
\includegraphics[width=7.5cm]{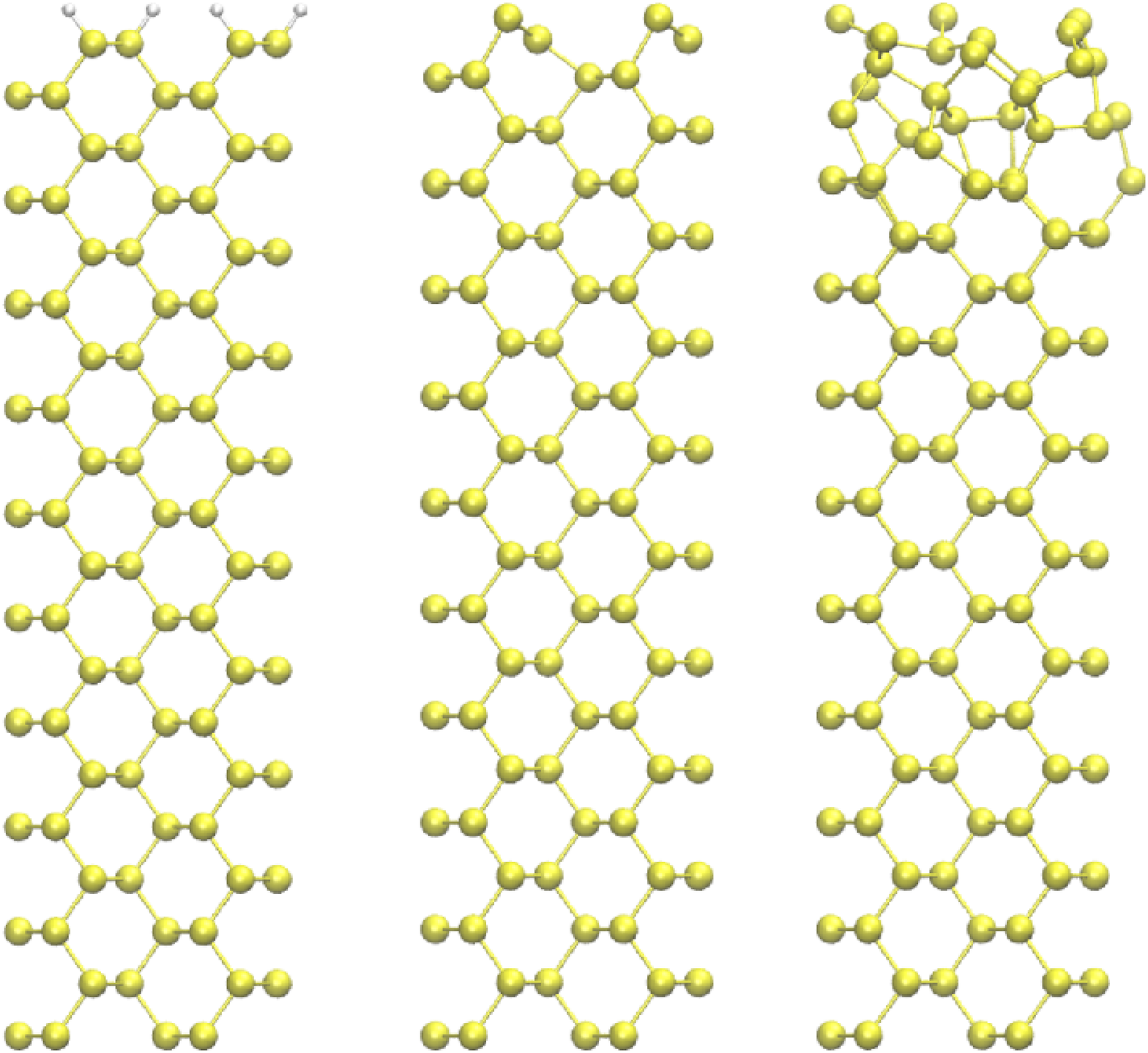}
\caption{Supercells used to calculate the out-of-plane strain of the Si(110) surface. Left: hydrogenated surface; center: reconstructed p(1$\times$1) surface; right: amorphous surface.} \label{110-supercell}
\end{figure}

\begin{figure}\centering
\includegraphics[width=7.5cm]{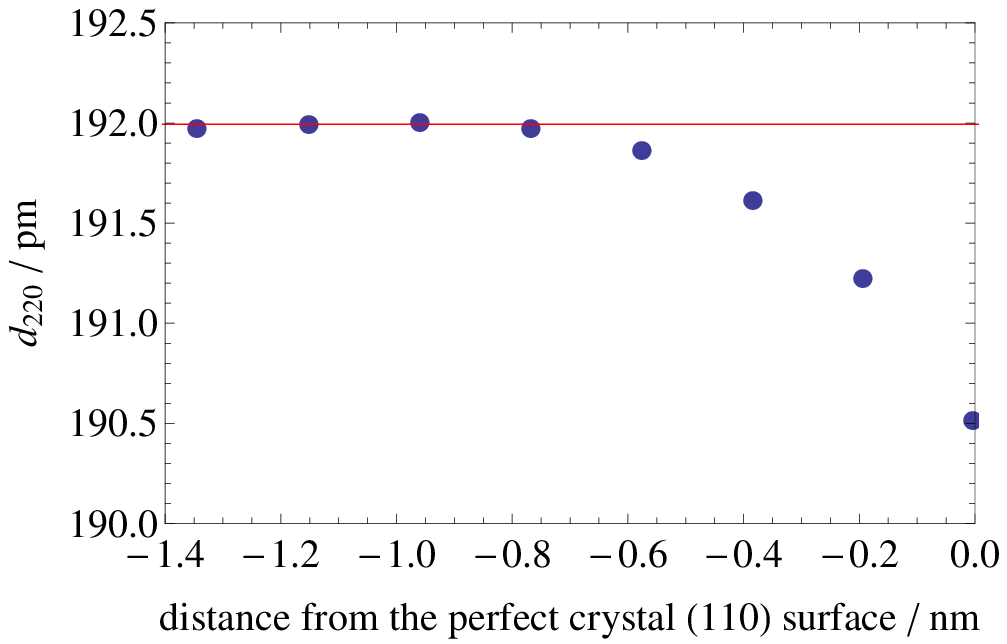}
\includegraphics[width=7.5cm]{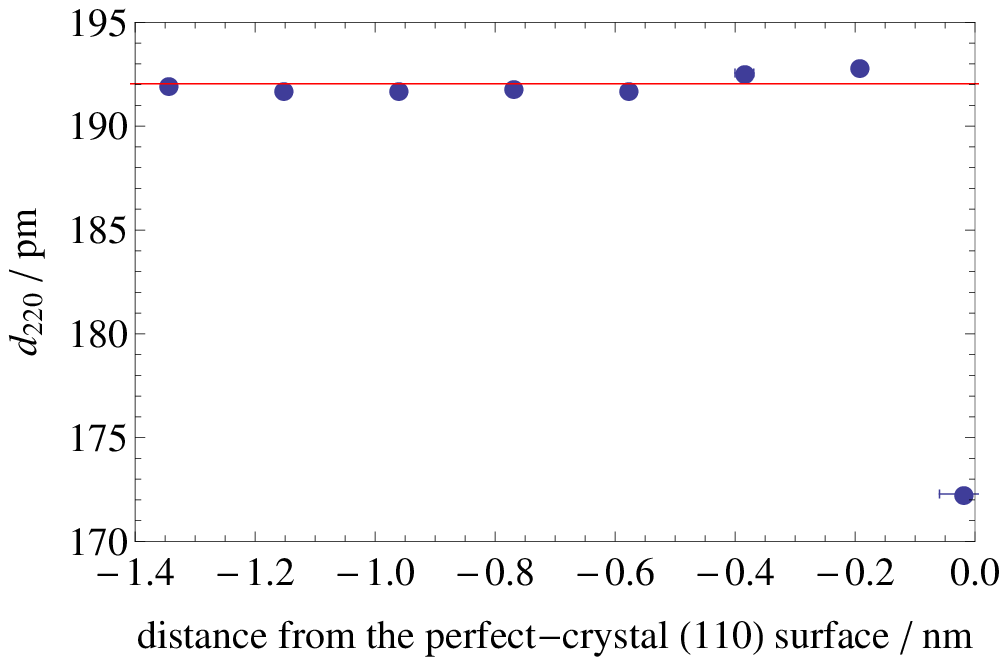}
\includegraphics[width=7.5cm]{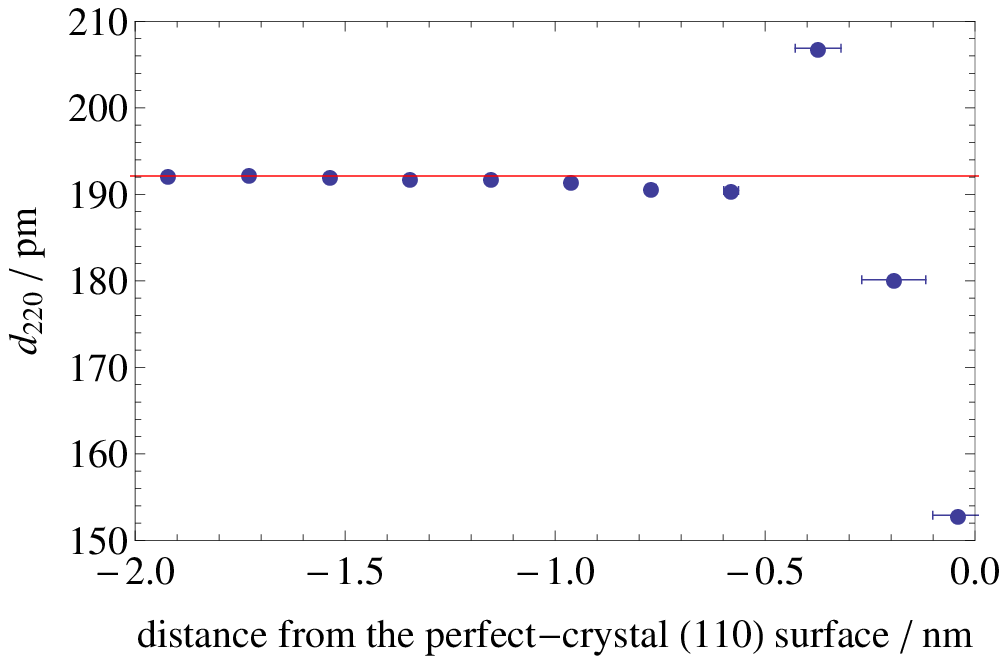}
\caption{Spacing $d_{220}$ of the \{220\} lattice planes as a function of the distance from the (110) surface of a perfect crystal. Top: hydrogenated surface; middle: p(1$\times$1) reconstruction; bottom: amorphous surface. The depth of each atom layer (blue dots) is calculated by sorting the supercell atoms by their distance from the surface and by taking the average of each subsequent subsets of 8 atoms. The error bars indicate the minimum and maximum depth of the atoms in each subset. The horizontal (red) lines are the perfect-crystal value of $d_{220}$.} \label{110}
\end{figure}

The (110) surface is of a particular interest, because it is also the surface of the lamellae of the $\eSi$ interferometer used to measure the lattice parameter. We investigated both the hydrogenated and reconstructed surfaces (Fig.\ \ref{110-supercell}, left and center). In the case of the hydrogenated surface, we observed a tiny spacing-variation of the \{220\} planes (up to about 0.7\%) involving the Si layers up to a deep of $-0.6$ nm from the surface (Fig.\ \ref{110}-top). Interestingly, the absolute variation is smaller than in the hydrogenated (100) surface (Fig.\ \ref{100}-middle).

\newpage

In addition, we investigated the clean surface, where a full structural optimization was allowed. The reconstruction of the (110) surface is still under debate, a clear picture of the reconstruction mechanisms is lacking, and several models have been proposed corresponding to different reconstructions \cite{110-recon}. During the geometry optimization we observed a $1\times 1$ relaxation of the surface which is fully consistent with previous works \cite{110-recon}. Figure \ref{110}-middle shows the $d_{220}$ spacing of the \{220\} planes as a function of their distance from the surface. We observe a large spacing variation (larger than 10\%) of the last two planes; therefore, we identify the last three atom layers as the surface region.

Ion-beam figuring of the ball and interferometer is being considered, which will creates an amorphous layer on the machined surfaces \cite{Paetzel:2014}. Although it can be removed via a sequence of natural oxidation and wet etching by HF, it is interesting to investigate what strain could be expected. Therefore, we carried out a combination of classical molecular dynamics and first principles DFT calculations to estimate the out-of plane lattice constant variations also in this case. In order to create an amorphous surface-layer we used a standard quenching-from-the-melt procedure and carried out a classical molecular dynamics simulation by means of the LAMMPS package \cite{LAMMPS}. The interatomic interactions have been sampled using the EDIP model potential \cite{EDIP}. The simulation protocol involved an initial annealing of the sample at 800 K for 5 ns; next we cooled down the sample for 5 ns at 300 K. The sample so obtained was further relaxed by a DFT simulation using the same parameters described above; Fig.\ \ref{110-supercell}-right shows the final structure obtained. Figure \ref{110}-bottom shows the $d_{220}$ variations; also in this case we observe a large variation (about 21\%) for the last layers.

\begin{figure}\centering
\includegraphics[width=7.5cm]{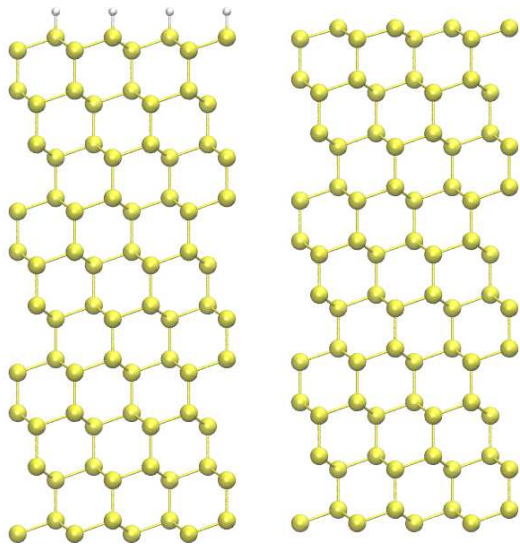}
\caption{Supercells used to calculate the out-of-plane strain of the Si(111) surface. Left: hydrogenated surface; right: non reconstructed and non saturated surface.} \label{111-supercell}
\end{figure}

\begin{figure}\centering
\includegraphics[width=7.5cm]{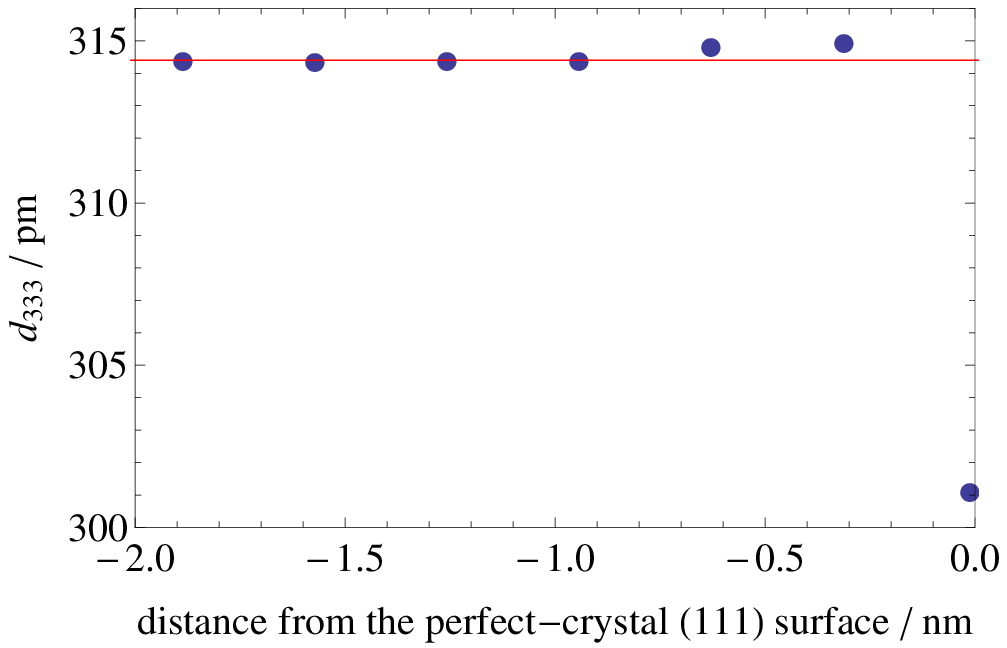}
\includegraphics[width=7.5cm]{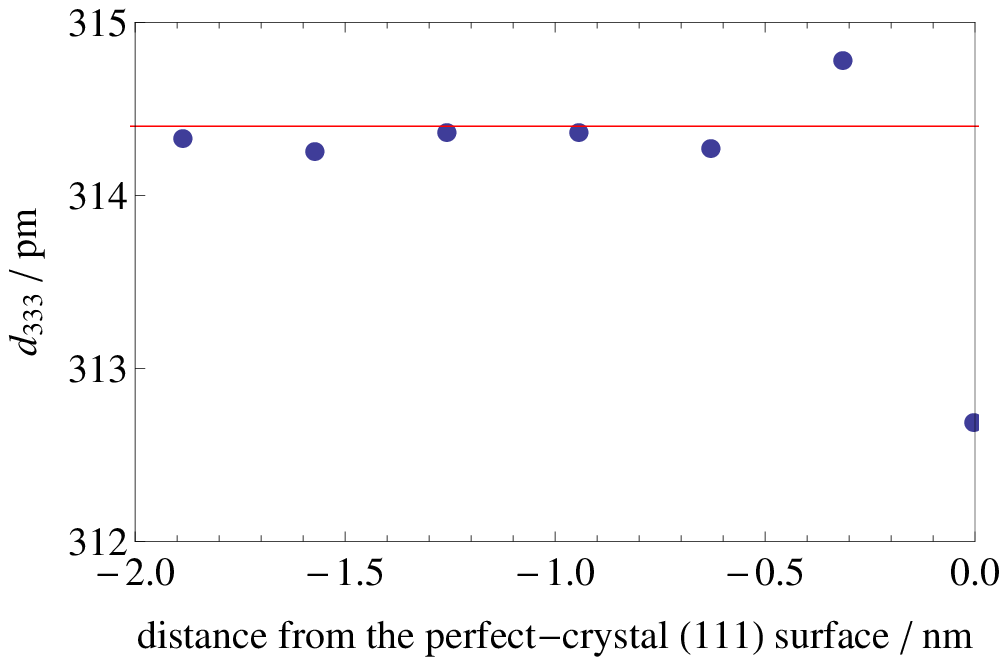}
\caption{Spacing $d_{333}$ of the \{333\} lattice planes as a function of the distance from the (111) surface of a perfect crystal. Top: non reconstructed and non saturated surface; bottom: non reconstructed and hydrogenated surface. The depth of each atom layer (blue dots) is calculated by sorting the supercell atoms by their distance from the surface and by taking the average of each subsequent subsets of 8 atoms. The error bars indicate the minimum and maximum depth of the atoms of each subset. The horizontal (red) lines are the perfect-crystal value of $d_{333}$.} \label{111}
\end{figure}

\subsection{(111) surface}
We completed the silicon surface strain characterization by considering the (111) surface. Differently from the (100) and (110) orientations, the (111) surface reconstruction has been fully experimentally characterized \cite{111-recon}. The most energetically stable reconstruction is a complex 7$\times$7 one having a large unit cell of 49 Si atoms. Unfortunately, a full first-principles characterization of such a complicated reconstruction, performed in a simulation cell having at least 20 Si layers, would result in a number of atoms (980) exceeding our computational budget. For this reason, we excluded the characterization of the (111) reconstruction by focusing only on the case of fully hydrogenated  and non-hydrogenated unreconstructed surfaces (see Fig. \ref{111-supercell}).\\ Figure \ref{111} shows the $d_{333}$ spacing of the \{333\} planes as a function of their distance from the surface. In the case of the non-hydrogenated unreconstructed surface (Fig. \ref{111}-top), we observe a sizable spacing variation, about 4\%, corresponding to the last silicon plane. Such a variation results to be greater than the one of the (100) surface and smaller than the the one of the (110) case. As far as concerns the hydrogenated surface (Fig. \ref{111}-bottom), we observe negligible spacing variations with respect to the perfect crystal.

\section{Correction of the volume measurement}
To estimate the effect of the surface on the measurement of a Si-ball volume, we calculated the difference between the radius of a perfect-crystal ball and the radius of a surface-strained one. To this end, we used the supercells of section \ref{CLS} as core-samples to extract information about the inner stratigraphy of the ball. We identified $d_{\rm hlm}^0=a_0/\sqrt{h^2+l^2+m^2}$, where $a_0=543.041$ pm is the Quantum Expresso value of the c-Si lattice parameter, with the perfect-crystal value of the spacings of the planes \{400\} -- (100) surface, \{220\} -- (110) surface, and \{333\} -- (111) surface. The perfect-crystal thickness of the 20 supercell layers was obtained as $T_0 = 19 d_{\rm hlm}^0$; in this way, we eliminated any bias due to the difference between the $a_0$ values provided by Quantum Expresso code and measurement.

The thickness of the outermost 20 perfect-crystal layers was compared against the thickness
\begin{equation}\label{TS}
 T_{\rm s} = \sum_{i=2}^{20} d_{\rm hlm}(i)
\end{equation}
of the same layers strained by the surface reconstruction and amorphous layer. The table \ref{T1} summarizes the results for the (100), (110), and (111) surfaces. In the amorphous-layer case, the layer thickness was calculated by averaging the vertical position of the 8 topmost atoms. In all cases, the thickness of the surface-strained supercell is smaller than the thickness of the same supercell in the crystal bulk.

In the worst case (amorphous-layer), the difference between the measured volume of a surface-strained ball and the volume of the same perfect-crystal ball is
\begin{equation}\label{V-err}
 \frac{\Delta V}{V} = \frac{3\Delta R}{R} \approx 2.5\times 10^{-9} ,
\end{equation}
where $R\approx 47$ mm is the ball mean-radius and $\Delta R=T_0 - T_{\rm s}=0.039$ nm, the measured volume being smaller than the volume assumed in (\ref{NA}).

\begin{table}[t]
\caption{\label{T1}Sinking of c-Si surfaces under the effect of surface reconstruction and amorphous or oxide layers.}
\begin{indented}
\item[]\begin{tabular}{lr}
\br
Surface &sinking / nm\\
\mr
Si(100) non-reconstructed and non-saturated &0.002\\
Si(100) non-reconstructed and hydrogenated  &0.005\\
Si(100) $2\times 2$ reconstruction &0.023\\
Si(110) non-reconstructed and hydrogenated  &0.003\\
Si(110) $1\times 1$ reconstruction &0.018\\
Si(110) amorphous layer            &0.039\\
Si(111) non-reconstructed and non-saturated &0.012\\
Si(111) non-reconstructed and hydrogenated  &0.003\\
\br
\end{tabular}
\end{indented}
\end{table}

\section{Conclusions}
According to the results of our simulation of the surface-induced strain, in the worst case, the volume of the $\eSi$ balls used to determine the Avogadro constant is smaller than the perfect-crystal volume in the measurement equation (\ref{NA}) by about $2.5\times 10^{-9} V$. This result relies on first-principles density-functional calculations of the out of plane lattice-parameter variation for the Si (100), (110), and (111) surfaces. Surface reconstruction and the presence of an amorphous layer were considered. The difference is an order of magnitude smaller than the present measurement uncertainty, about $2\times 10^{-8} V$, but it is not as small as it might be expected and it may be worth to be considered in future, more accurate, measurements.

\section*{Acknowledgements}
This work was jointly funded by the European Metrology Research Programme (EMRP) participating countries within the European Association of National Metrology Institutes (EURAMET) and the European Union. C.M. acknowledges Sardinia Regional Government for financial support (P.O.R. Sardegna ESF 2007-13). G.M. thanks Petr Kren of the Czech Metrology Institute for having brought to our attention the surface stress problem.

\end{document}